# Low frequency noise statistics for the breakdown characterization of ultra-thin gate oxides

N. Z. Butt
*School of Electrical & Computer Engineering, Purdue University, West Lafayette, Indiana 47907*

A. M. Chang[a]
*Department of Physics, Purdue University, West Lafayette, IN 47907-1396, and Department of Physics, Duke University, Durham, North Carolina 27708-0305*

H. Raza and R. Bashir
*Birck Nanotechnology Center and School of Electrical & Computer Engineering, Purdue University, West Lafayette, Indiana 47907*

J. Liu and D. L. Kwong
*Electrical & Computer Engineering, University of Texas at Austin, Austin, Texas 78712-0240*

We have investigated the *statistics* of low frequency noise in the tunneling current of ultrathin oxides (2.5nm-4nm) in metal oxide semiconductor capacitors as a function of the applied voltage stress. The statistical analysis includes (i) non-Gaussianity (nG), which is a measure of the degree of temporal correlation in the noise, and (ii) ratio of integrated noise power to the DC leakage current (R). The occurrence of high peaks in nG indicates the appearance of new percolation paths, and the subsequent conduction through these paths is indicated by R. Our results show that the nG and R characteristics are generic for the oxides of different thickness and growth quality and have the potential, in conjunction with leakage itself, of being used as a prognosticator of oxide reliability.

There has been a growing desire to precisely characterize the process of oxide breakdown to have an accurate estimation for the reliability of ultrathin gate oxides of MOS transistors. When a voltage stress is applied across an ultrathin capacitor, the conduction through it can be categorized as of two types. One is the direct tunneling of carriers through the oxide barrier and the other is the tunneling through the defects (traps) present in the oxide, which is called the trap-assisted tunneling (TAT). With the increase of stress, the density of traps increases and new conduction paths are formed inside the oxide, giving rise to breakdown events. In thinner oxides, the characterization becomes difficult because the breakdown events are often very soft and are not easily detected in the (commonly measured) average leakage current density. On the contrary, it was earlier reported that the low frequency 1/f noise in the tunneling current can be very sensitive to the damage associated with the trap assisted processes [1,2] and it was observed that the emergence of transient current spikes in TAT made the noise highly non-Gaussian. In our work, we have studied the characteristics of noise in $SiO_2$ capacitors by applying a ramp of voltage stress ($V_g$) at the gate in small incremental steps: these capacitors have an area $\sim 10^{-4} cm^2$ and differ for the oxide thickness and the growth process. Analyzing the statistics of noise in each Vg step, we found interesting signatures in nG and R, which are supposed to be closely related to the creation and development of percolation paths leading to hard and soft breakdowns (HBD and SBD respectively). In contrast, no notable features, in either nG or R, were observed in a 100 MΩ resistor or in any capacitor after it had undergone a HBD. These signatures of breakdown pathways were similar in oxides of different thickness

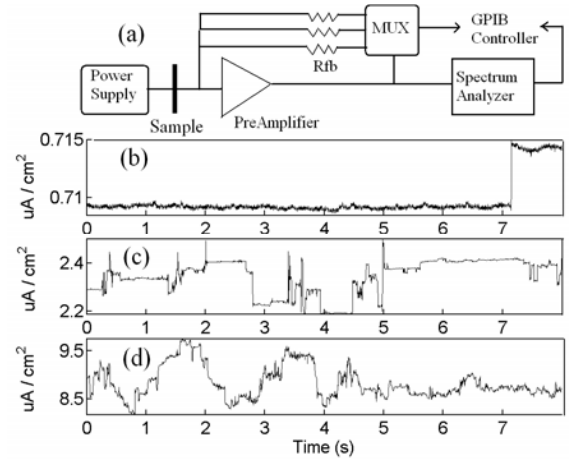

Fig. 1 (a) The experimental setup. (b-d) Examples of time traces of leakage current density ($J_{dc}$) illustrating the sequence of a progressive BD in a 4nm oxide. Traces (b-d) were measured at incremental Vg values in the range spread by a big R peak (b) appearance of sharp current spike which resulted in high nG in noise before any increase in R (c) sharp increase in noise with a relatively small increase in $J_{dc}$ which resulted in a big jump in R (d) increase in $J_{dc}$ due to added SILC contribution from the BD with no further increase in noise; this resulted in the falling of a R peak.

---

[a] Electronic mail: yingshe@phy.duke.edu

and growth, although the magnitude of the stress needed to observe them was higher for good quality devices. The following three phenomena, which occurred sequentially, were involved in the breakdown (BD) pathway occurring at a certain $V_g$: (i) the appearance of nG peaks in the noise (ii) the jumps in R and (iii) an increase in the stress induced leakage current (SILC). The very first indication that a BD might occur was the presence of large peaks in nG, which were a result of the appearance of sharp transitory spikes in TAT. This was due to the creation of transitory conduction paths, in the oxide layer, from newly generated traps. The subsequent development of conduction through these transitory paths was manifested in the corresponding values of R as a function of $V_g$. In a breakdown pathway, peaks, due to the increased fluctuations (noise) in TAT, were observed in R - while there was no significant increase in the DC leakage current density $J_{dc}$: These fluctuations were presumably a direct consequence of conduction through the developing percolation paths through the oxide. Additionally it was found that a correlation existed between the magnitude and the spread (over $V_g$) of the R peaks on one side and the damage in the device, detected by the increased values of SILC at the corresponding $V_g$, on the other.

In particular, the magnitude of an R peak correlated to the severity of BD and its spread over Vg, to the progressiveness of BD. Based on these metrics, we found three types of breakdown in devices according to their severity and progressiveness. Type I was those SBDs characterized by small narrow jumps in R. These were not progressive and had a mild impact on damage depicted in SILC. Type II was SBDs which showed big peaks in R spread over a wide range of Vg indicating the progressive damage to the percolation path. This type of BD had a significant impact on SILC. Type III was HBDs which were indicated by narrow but very sharp increase in both R and SILC. These were rare in occurrence even in thicker oxides but were most severe to the oxide damage. The noise statistics based on nG and R can thus be a very useful probe of different types of BD events in oxides and can be used as a complement to the high sensitivity I-V measurements in order to precisely predict the reliability of ultrathin oxides.

An example of time traces illustrating the sequence of a progressive (type II) breakdown is shown in Fig. 1. The sharp jump of a relatively low magnitude in $J_{dc}$ in fig. 1(b) is a type of transitional current spikes, which resulted in high nG in noise. Until this point, there was no indication of breakdown in R or in $J_{dc}$. Fig. 1(c) shows the onslaught of spikes in TAT which indicated the start of a very noisy conduction presumably through a single percolating path. The increase of noise was large compared to the increase in $J_{dc}$ which resulted in a big jump in R. The trace in Fig. 1(d) indicates conduction through the percolation path during the final stage of BD when noise stopped increasing and $J_{dc}$ had a significant increase due to added SILC contribution from the BD.

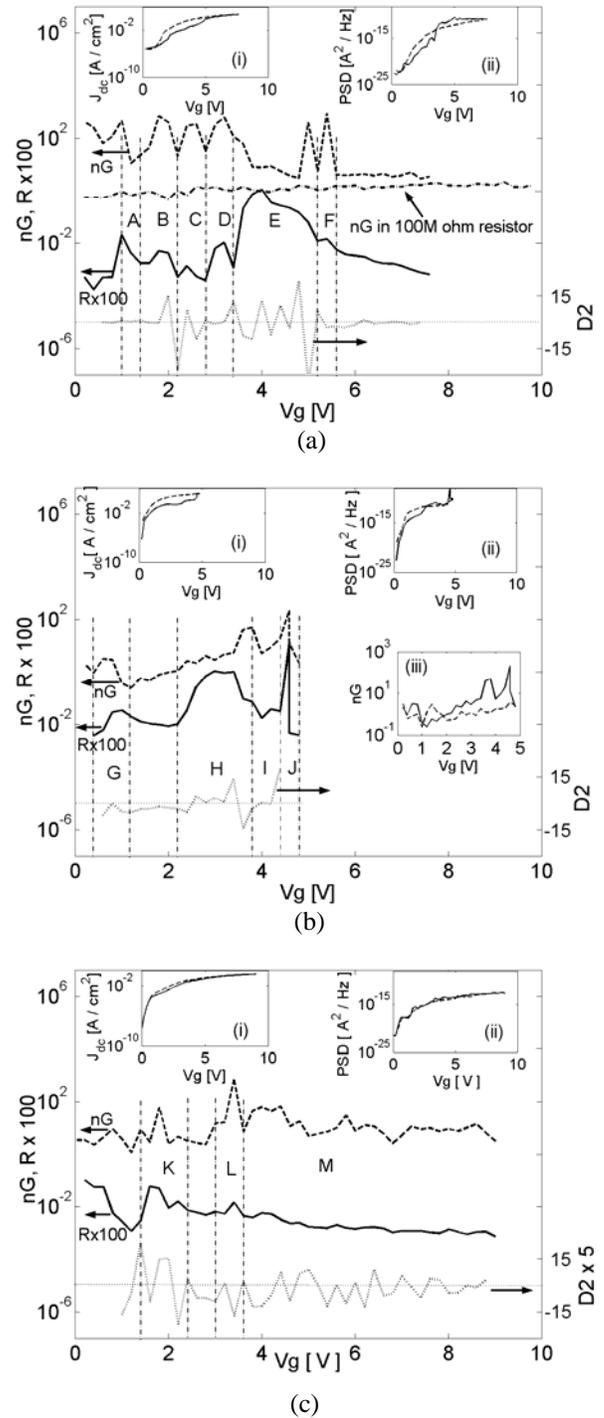

Fig. 2 The nG, R and D2 characteristics during the first run of voltage stress for (2a) 4nm device (2b) 3.35nm device (2c) 2.5nm device. The solid and the dashed styles represent R x 100 and nG respectively both as a function of $V_g$. The dash-dot curve in fig. 2a is the nG characteristics of a 100 MΩ resistor measured by applying a run of $V_g$ stress. The insets (i) and (ii) show $J_{dc}$ vs. $V_g$ and PSDI vs. $V_g$ characteristics respectively for two consecutive run of stress. The inset (iii) in fig. 2b shows the nG in two consecutive runs of stress to the 3.35nm device. The solid and the dashed styles represent the first and the second runs respectively in all the insets. The polarity of the voltage was negative at the gate for all data presented in this study. The vertical columns labeled as A, B, C etc. indicate different regions of breakdowns during the $V_g$ stress.

We have measured different quality oxides which differed for their thickness (in the range 2.5-4 nm) and their growth conditions. The 2.5nm samples were fabricated by Texas Instruments. The 4 nm oxides were grown in Purdue by dry oxidation for 30 min at 750 $^o$C at atmospheric pressure. The substrates were p-type silicon with boron doping of ~ $10^{15}$ cm$^{-3}$ and the gates were fabricated in aluminum deposited by e-beam evaporation. The capacitors were patterned and etched with different areas ranging from 78 x $10^{-6}$ cm$^2$ to 650 x $10^{-6}$ cm$^2$. The 3.35 nm samples were provided by Bell-Laboratories and were grown by rapid thermal annealing. The presence of a large density of native traps in the oxide was already known in this growth [3]. The thicknesses were verified by ellipsometry and cross-section TEM. The experimental setup is shown in Fig. 1(a) and consisted of a variable power supply, a low noise preamplifier to measure the current and a Stanford Research Systems SR785 spectrum analyzer (fig. 1a).

The noise characteristics were observed by applying a negative voltage stress at the gate in incremental steps of 0.2V. At each of these $V_g$ steps, about 300 traces (8 sec. each) of the gate leakage current were taken. The time traces were Fourier transformed after subtracting the DC current ($I_{dc}$) and the average power spectral density thus calculated was then integrated through the total bandwidth to obtain the integrated power spectral density (PSDI). The background noise was found to be a combination of the current and voltage noise of the operational-amplifier and was subtracted from the 1/f noise. The sensitivity of the measurements was down to 4 fA /$\sqrt{Hz}$ at 10Hz. R was calculated by the ratio: R=PSDI / $I_{dc}^2$. The non-Gaussianity of noise was computed at each of the constant voltage step using the expression [1]: nG= $<(S_i-S_{av})^2>n_b / S_{av}^2$, where $S_i$ is the average noise power in a single Fourier transform, $S_{av}$ is the average noise power for a set of N measurements, $n_b$, is the total number of bins in the bandwidth, and the brackets < > denote averaging over the set of N measurements. This quantity, nG, is simply a normalized variance of a series of noise power measurements and is a statistical test very sensitive to the transients in the time traces [1]. For a 1/f signal, without any transients, the noise was Gaussian with nG ~1 as can be seen from the results for a 100 MΩ resistor in Fig. 2a.

For each of the three types of devices, an example of nG and R vs. $V_g$ is shown in Fig. 2(a-c) for the first run of voltage stress. The 2$^{nd}$ derivative of log ($I_{dc}$) normalized by the |log($I_{dc}$)| is also plotted (D2): the jumps in D2 represent the damage to the oxide showed up in the local increase of SILC. In Fig. 2, the vertical lines separate the regions of BDs of different types (A, B, C, …, M). Large nG peaks started to appear in the 4nm device shown in Fig. 2a from very low $V_g$ and the small peaks in R illustrated type I breakdown (regions A-D). A type II breakdown initiated with a big jump in R around 4V (region E). The wide spread of R indicated a progressive damage, during which, SILC increased gradually as also indicated by a number of D2 jumps. The 3.35nm device in Fig. 2b showed a small peak in nG and R in region 'G' after which R became steady until a type II BD showed up around 2V (region H). It should be noted that due to a large density of native traps in 3.35nm growth, this device had an excessive leakage density even at very low $V_g$ (fig. 2b, inset (i)). The averaging effect of large number of readily available paths might be the reason that no sharp jumps in nG were observed in this device. Rather a steady rise of nG was observed with increasing $V_g$ which might be an indication that some particular transitory paths were gradually dominating over others. Finally this device showed a type III BD illustrated by a big sharp R jump as well as a jump in the SILC at approximately 5V (Fig 2b and inset (i)). The characteristics of 2.5nm device which are shown in Fig. 2c illustrate type I breakdowns in a 2.5nm oxide. The jumps in R and D2 were smaller and no type II or type III breakdowns were observed. The relatively significant jumps were found in region 'K' which was followed by a number of small jumps in regions 'L' and 'M'. The absence of more severe BD types (I and II) could be due to either or both of the following reasons: (i) the presence of very few native traps due to a better good growth quality (ii) the lesser oxide thickness might had an effect to reduce the probability of both progressive (Type II) and hard (Type III) BDs.

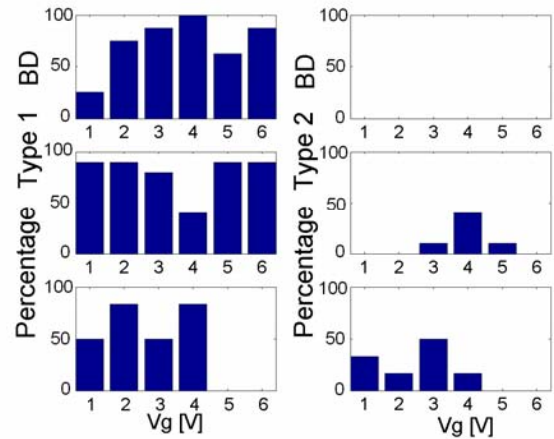

Fig. 3 The statistics of Type 1 and Type 2 BD occurrences at different magnitudes of applied stress in the data collected on 6-10 devices of each type. From Top to Bottom: 2.5 nm, 4nm and 3.35nm respectively.

The statistics of Type I and Type II breakdowns, in three different oxide types, are shown in Fig. 3. Type III breakdown is not shown since it was observed not more than one of the 3.35nm devices. These statistics represent the percentage breakdowns in the each of the three oxide types as a function of $V_g$. The probabilities for BD correlate well with the device quality, where the better quality devices show a reduced tendency for BD at a given stressing voltage.

Aside from the ability to distinguish between three types of qualitatively different BD processes, the most notable finding of our noise characterization is the observation that while the pathway to BD qualitatively appears remarkably similar for oxides of differing quality, the noise signatures differ quantitatively. This bodes well of further development of the noise characterization method as a sensitive tool to ascertain oxide quality and to study the mechanism of BD, e.g. in order to distinguish between different models of SBD [4-6].

The authors would like to acknowledge Tom Sorsch for providing 3.35 bell-lab samples. This work was supported by NSF ECS 0100202.